\documentclass[aps,twocolumn,pra,showpacs,amsmath,nofootinbib]{revtex4}

\usepackage{graphicx}
\usepackage{dcolumn}
\usepackage{bm}
\usepackage{epstopdf}
\usepackage{multirow}
\usepackage{wrapfig}
\newcommand{\beq}{\begin{equation}}

\newcommand{\eeq}{\end{equation}}
\newcommand{\bea}{\begin{eqnarray}}
\newcommand{\eea}{\end{eqnarray}}
\newcommand{\eps}{\varepsilon}
\newcommand{\frc}[2]{\raisebox{1pt}{$#1$}\big/\raisebox{-1pt}{$#2$}}

\begin{document}

\author{S. Kamerdzhiev\footnote{kamerdzhievsp@nrcki.ru}}
\affiliation{National Research Center Kurchatov Institute, Moscow,
123182 Russia.}

\author{M. Shitov}
\affiliation{National Research Center Kurchatov Institute, Moscow,
123182 Russia.}

\author{D. Voitenkov}
\affiliation{Science and Innovation  Joint-Stock Company, Moscow, 119180 Russia.}

\title{Microscopic theory
of pygmy- and giant resonances: \\ Accounting for complex 1p1h$\otimes$phonon
and two-phonon  configurations }


\begin{abstract}
The self-consistent  Theory of Finite Fermi Systems (TFFS) is consistently generalized for   the case of accounting for phonon coupling (PC) effects 
in  the energy region of pygmy- and giant multipole resonances (PDR and GMR) in magic nuclei with the aim to consider particle-hole (ph) and both complex
1p1h$\otimes$phonon and  two-phonon configurations. The article is the direct continuation and generalization  of the previous article 
[S.Kamerdzhiev, M.Shitov, Eur.Phys.J.A. \textbf{56}, 265 (2020)],  referred to as [I], where 
1p1h- and only complex 1p1h$\otimes$phonon configurations were considered.
  The newest equation for the TFFS main quantity,  the effective field (vertex), which  describes the nuclear polarizability, has been obtained.
 It has considerably generalized the results of the previous article
 and accounts for two-phonon configurations. 
Two variants of the newest vertex equation   
have been derived: (1)the first variant contains complex 1p1h$\otimes$phonon configurations and the full 1p1h-interaction amplitude  $\Gamma$ 
instead of the known  effective interaction F in [I],
 (2) the second one contains both 1p1h$\otimes$phonon and   two-phonon  configurations. Both variants contain  new, as compared to usual approaches, PC contributions,    
  which are of interest in the energy region under consideration and,  at least, should result in a redistribution of the PDR and GMR strength,
which is important for the explanation of the PDR and GMR fine structure. 
   The qualitative analysis and discussion  of the new terms and the comparison to the known time-blocking approximation   are performed.

\end{abstract}

\maketitle

\section{Introduction}
The experimental and theoretical studies of multipole pygmy-and giant resonances, especially pygmy-dipole resonance  (for simplicity, hereinafter  PDR and GMR), continue to draw  much attention, see reviews \cite{Savran,Paar,Bracco,revKST,kaevYadFiz2019}. For PDR , it is explained by the new    experimental possibilities \cite{Cosel,Larsen, Bracco}, for example, polarized proton inelastic scattering at very forward angles \cite{Cosel},  the existence of many new and delicate physical effects in this energy region, like irrotational and  vortical kind of   motion 
 \cite{Nester, Paar}
and the  upbend phenomenon for the photon strength function  in the energy region of 1-3 MeV  \cite{Larsen}. 
Besides, as it turned out, it is impossible to explain completely, with the account of phonon coupling (PC), the observed PDR fine structure,  even in the nucleus $^{208}$Pb, within both non-self-consistent  
\cite{rezaeva} and self-consistent \cite{Lyutor2018} approaches.

From a physical point of view, the problem is understandable, but only in principle: it is necessary to take into account the phonon coupling (PC) effects in addition to the standard random phase approximation (RPA) and quasiparticle RPA (QRPA) approaches. Here, several non-self-consistent and self-consistent approaches have been developed, see reviews \cite{Paar, Bracco,kaevYadFiz2019}, 
which considered  both  1p1h$\otimes$phonon and
1p1h$\otimes$phonon + two-phonon  configurations. But they still have room for the improvement.  
We are sure that  the use of  consistent quantum many-body formalism,  to be more exact, the Green function (GF) method, including  generalization of the self-consistent Theory of Finite Fermi Systems (TFFS) \cite{Migdal,KhSap1982} is very promising for further work.

In the self-consistent TFFS, which was developed in \cite{KhSap1982} and partly described in the second edition of  Migdal's book
\cite{Migdal2}, it was shown that in all  the numerous  problems considered \cite{kaevYadFiz2019} PC contributions were considerable,  of fundamental importance and  necessary for explaining experimental data. In the works \cite{KhSap1982} only  the self-consistent description  of characteristics of the ground and low-lying collective states was considered.
The success of these works is explained, first of all,
 by a more consistent consideration of  PC effects, in particular, the tadpole effects \cite{KhSap1982,SapTol2016} and, probably, by the effect of  the effective interaction variation  in the phonon field \cite{voitenkov}.

  In our opinion, the next natural step is to go further and to consider the field of PDR + GMR within the self-consistent TFFS. 
This is the general aim of our article

  The theory for PDR and GMR was   developed, within the GF method and with PC,  in the framework of both non-self-consistent \cite{kaev83,ts89, revKST} and self-consistent variants      \cite{ts2007,ave2011}. The  difference between \cite{kaev83} and \cite{ts89} consisted in the fact  that in \cite{ts89} the disadvantage of \cite{kaev83} was
eliminated, namely, a special approximate method 
    of chronological decoupling of diagrams (MCDD) (or, in a more modern terminology, time blocking approximation (TBA)) was developed in order to solve the problem of  second-order poles in the generalized propagator of the extended RPA propagator in \cite{kaev83}. This disadvantage was not important  for explanation of  the properties of M1 resonance \cite{ktZPhys,kt1984}, which is in the PDR energy region. Moreover, earlier  it was even used within Nuclear Field Theory (NFT)  for 
electrical GMR \cite{Bortignon,Bortignon2} with the smearing parameter of 600 keV.
Later on, this method to solve  the second-order poles problem was considerably improved \cite{ts2007} so that the approach  obtained the name of  quasiparticle time blocking  approximation (QTBA) for nuclei with pairing and TBA for magic nuclei, respectively. TBA and QTBA and their  modifications  have been  applied for the description of PDR and also GMR in magic and semi-magic nuclei \cite{ts2018,litva-schuck,Larsen,kaev2014}.  Usually, in the most part of  these works the calculations were performed with the use of a smearing parameter, which was  taken to be equal to the experimental resolution of about 100-500 keV. However,  the main physical content, i.e. inclusion of PC \textit{only} into the   particle-hole (ph) propagator 
    (in the language of TFFS) was always preserved  despite the fact that the derivation method  was different and was based on the Bethe-Salpeter equation.   
 Unfortunately,  even the known  tadpole effect
 was not considered in the generalized  propagator of  MCDD, or TBA.
 
     In  article \cite{KaShi}, which further will be referred  to as [I], a theory of PDR and GMR  was  developed within the GF method in the  approximation 
of 1p1h+1p1h$\otimes$phonon configurations. The consistent generalization of the standard TFFS with the aim to include  $g^2$ PC corrections, where $g$ is the  creation amplitude of the usual RPA phonon, allowed us to obtain  new PC contributions to the new equation for the main TFFS quantity - vertex V. They included the 
dynamical  tadpole effects for the vertex, the so-called induced interactions caused by the phonon exchange and the first and second variations 
of the effective interaction in the phonon field. In article [I], we  used the approximation for the first and second variations of the vertex V in the phonon field, namely, only  free terms of the  equations for them were used, which provided accounting for only complex
 1p1h$\otimes$phonon configurations. In the present article, we reject   this approximation and consider
 the exact equations for the vertex variations, i.e. accurate expressions obtained for them.  

The restriction by only complex 1p1h$\otimes$phonon configurations is not enough in the PDR+GMR field, at least for the explanation of the PDR fine structure.  In the non-self-consistent 
approaches, this can be seen from  numerous calculations in the PDR energy region, see \cite{Savran,rezaeva}, performed within the non-self-consistent Quasiparticle-Phonon Model (QPM) \cite{Sol89}. 
 In order to improve the agreement with the 
$(e,e^\prime)$-experiments there, it was necessary to add more complex, first of all, two-phonon configurations. As it was said earlier, the comparison to the latest  experimental 
results obtained with polarized proton inelastic scattering at very forward angles \cite{Cosel}, also confirmed that, within 
the self-consistent approaches with only 1p1h$\otimes$phonon configurations \cite{Lyutor2018}, it was not possible to explain the PDR fine structure in $^{208}$Pb. 
The importance of the role of two-phonon configurations in this problem was also underlined  in self-consistent calculations of \cite{it-z2015}.

The goal of  this article is to do the  next necessary and essential step further than in  article [I]. 
Namely, to analyse the consequences of the use of accurate expressions for the first and second variations
of the vertex in the phonon field instead of the approximate ones, and, as one of these consequences, to add 
the second type of complex configurations, i.e., two-phonon ones.

In this article, we will consider only  magic nuclei,  complex  1p1h$\otimes$phonon and  two-phonon configurations.
As  usual, we use the fact of existence of
the small $g^2$ parameter. Very often we symbolically write  our formulas, the main of which are represented in the form of Feynman diagrams, so the final formulas can be easily obtained.

\section{Some earlier results}

\subsection{Some initial formulas of TFFS.}
In the standard TFFS, the main quantity in the problems connected with the interaction between a nucleus and an external field $V^{0}(\omega)$ 
with the energy $\omega$,  is the notion of
effective field (vertex) $V$, which    describes the nuclear polarizability and  satisfies the equation in the symbolic form \cite{Migdal}:
\begin{equation}
V = e_{q}V^0 + FAV,
\label{formula1}
\end{equation} 
where the ph-propagator reads: \footnote{We write down the standard TFFS
expression (2) for the ph-propagator as the result of integration. However, it is desirable to keep in mind 
that in  following Section IV it is better to consider the propagator as the product GG of two GFs without integration.} 
\begin{equation}
A_{12} (\omega) = \int G_1(\eps )G_2(\eps - \omega)d\eps. 
\label{formula2}
\end{equation} 

The full 1p1h-interaction amplitude $\Gamma (\omega)$ satisfies the equation:
\begin{equation}
\Gamma = F + FA\Gamma
\label{formula3}
\end{equation} 
In Eq.(\ref{formula1}) and Eq.(\ref{formula3}), F is the effective interaction, which in the self-consistent TFFS is calculated as the second variational derivative
of  the energy density functional and  the mean field is calculated as the first  variational derivative
of  the energy density functional. Low indices mean a set of single-particle quantum numbers 1 $\equiv ({n_1,j_1,l_1,m_1})\equiv \lambda_1$ and we write  $d\eps$ instead of $\frac{d\eps}{2\imath\pi}$ everywhere.  
 
The phonon  creation amplitude g satisfies the equation \cite{Migdal}  
\begin{equation}
g = FAg.
\label{formula4}
\end{equation} 
Eq.(\ref{formula1}), Eq.(\ref{formula3}) and  Eq.(\ref{formula4}) comply with  the RPA  approach written in the FG language,
i.e. with  the standard TFFS. We will use them as the initial relations or input data  for further development. And  we 
will speak about  generalization of the \textit{standard} TFFS just in this sense in the present article. 
  
\subsection{Earlier results with PC}
 
\begin{figure}[h]
\includegraphics[width=1\linewidth]{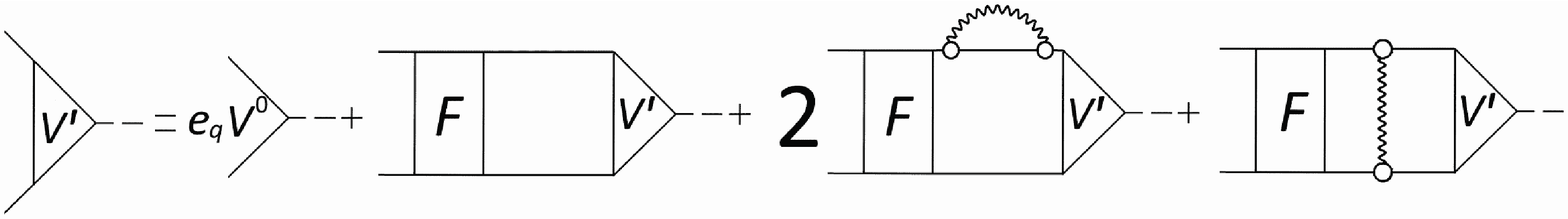}
\caption{Equation for the vertex $V^{\prime}$ containing the  simplest ph-propagator with PC \cite{kaev83,ktZPhys}.
Straight  and wavy lines correspond to GFs G and D, circles
with one wavy line stand for the amplitude of phonon production g. The rectangle stands for the effective interaction F. See text for details.}
\label{fig-1}       
\end{figure}

As it was mentioned in the Introduction, the physical content  of the previous GF approaches   in the PDR and GMR theory consisted in the fact  that  $g^2$ PC corrections were included only into ph-propagator, Eq.(\ref{formula2}). So, in the language of TFFS,  the diagrams   presented  in Fig.\ref{fig-1} should correspond to the  equation for the  vertex  $V^{\prime}$ with the simplest  PC ph-propagator. The case  without the MCDD prescription was realised in \cite{kaev83,kt1984,ktZPhys} for M1 resonances in magic nuclei. Physically, they also  correspond to the approach within NFT \cite{Bortignon,Bortignon2} for GMR. 
In Fig.1
 the  diagrams without phonons represent the   RPA case for the vertex V formulated in the standard  TFFS language, Eq.({\ref{formula1}), with the ph-propagator A, Eq.(\ref{formula2}).
Hereinafter, number  2 before a graph or a corresponding formula means that there  are two graphs or formulas of a similar type.  Of course, it is  necessary to keep in mind that  in reality  the  MCDD prescription \cite{ts89}, or TBA,   should be used for  numerical results to avoid the above-mentioned problem with  the second-order poles in the PC propagator shown in Fig.\ref{fig-1}. Such a generalized MCDD propagator is rather complicated, it   was discussed in details and given in \cite{revKST}. 

In the works performed within the self-consistent TFFS, the  $g^2$ PC corrections to the mean field, which take into account  the tadpole,   were  actively used.  They can be written in the symbolic form  as
\beq
\delta \Sigma = gDGg + g_{11}D
\label{formula5}
\eeq
\begin{figure}
\includegraphics[width=1\linewidth]{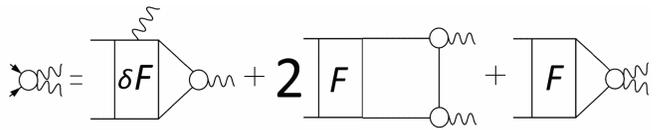}
\caption{Diagram representation of Eq.(\ref{formula7}).  The circle with two  wavy lines is $g_{12}$, which (with 1=2) is contained in the tadpole}
\label{fig-2}       
\end{figure}
Here $ \Sigma$ is the self-energy operator, G and D are the single-particle and  phonon  GF's:
\bea
G_1(\eps) = \frac{1-n_1}{\eps -\eps_1 + \imath\gamma} +  \frac{n_1}{\eps -\eps_1 - \imath\gamma},  \nonumber \\
D_s(\omega) = \frac{1}{\omega - \omega_s + \imath\gamma} - \frac{1}{\omega + \omega_s - \imath\gamma},  
\label{formula6}
\eea
  g obeys the homogeneous equation (\ref{formula4}) and $\eps_1 \equiv \eps_{\lambda_1},n_1 \equiv n_{\lambda_1} $

The amplitude of creation of two  phonons, similar  in the tadpole  case, is obtained as the variation of Eq.(\ref{formula4}) in  the field of  the phonon 2
\bea
g_{12} \equiv \delta^{(1)}g_2 = \delta^{(1)}FAg_2 + F\delta^{(1)}Ag_2 + FAg_{12}
\label{formula7}
\eea
This equation shown in Fig.\ref{fig-2} was solved only in the coordinate representation  in the works 
pertaining to other  problems connected with the properties of the ground or low-lying collective states \cite{platonov,KhSap1982}. In the works
of Kurchatov Institute group \cite{KhSap1982,SapTol2016},  a realistic estimation of the two-phonon creation amplitude $g_{11} \equiv \delta^{(1)}g_1$ contained in the phonon tadpole term $K^{ph} = \int d\omega g_{11}D$ was used. This estimation was  based on the ansatz 
for the quantity $\delta F$ \cite{KhSap1982}:
\beq 
\delta F = (\frc{\delta F}{\delta \rho})\, Ag
\label{formula8}
\eeq

\section{Exact expressions for first and second variations of the vertex $\delta V$ and $\delta^{(2)}V$}

In order to obtain the full $g^2$ corrections to the vertex V, Eq.(\ref{formula1}), we will use, like in [I], the following expressions for them
 \beq
\tilde V = V + \Delta V(g,  V)
\label{formula9}
\eeq
and 
 \beq
\Delta V = 2gDG\delta^{(1)}V + \delta^{(2)}VD,
\label{formula10}
\eeq
where the quantities $\delta^{(1)}V$ and $\delta^{(2)}V$ are  the first- and second-order variations  of the vertex $V$, Eq.(\ref{formula1}),
in the phonon field.
   These corrections  are  shown in Fig.\ref{fig-3}. 
   
 \begin{figure}
\includegraphics[width=1\linewidth]{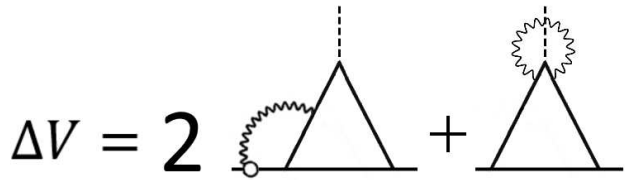}
\caption{The $g^2$ corrections to the vertex V, Eq.(\ref{formula10})}
\label{fig-3}       
\end{figure}  
    The second term in Eq.(\ref{formula10}), Fig.\ref{fig-3}, contains  "pure"  $g^2$ corrections,
while the first term  in Eq.(\ref{formula10}), Fig.\ref{fig-3}, is a mix of the first-order correction to the vertex $V$  and the "end" corrections of the
   first order in $g$.
   
     First of all, let us obtain the quantity 
     $\delta ^{(2)}A$ contained in $\delta^{(2)}V$ for our case of similar phonons. 
     In order not to confuse it with the single-particle index 1, here we   introduce the   notion $\tilde 1$
for the phonon 1.
      When $\tilde 1 = \tilde 2$, which is of interest in our case  of the variation
$\delta^{(2)}V = \delta^{\tilde 1}\delta^{\tilde 1}V$,  we obtain five terms (instead of eight in the general case of unequal 
      phonons for $g_{\tilde 1\tilde 2}$); they  are shown in Fig.\ref{fig-4}.
\bea
\delta^{(2)}A = \delta^{\tilde 1}\delta^{\tilde 1}G_1G_2 = 2G_1g_{\tilde 1}G_4g_{\tilde 1}G_3G_2 + \nonumber \\ 
2G_1g_{\tilde 1\tilde 1}G_3G_2 + G_1g_{\tilde 1}G_3G_2g_{\tilde 1}G_4
\label{formula11}
\eea

\begin{figure}[h]
\includegraphics[width=0.9\linewidth]{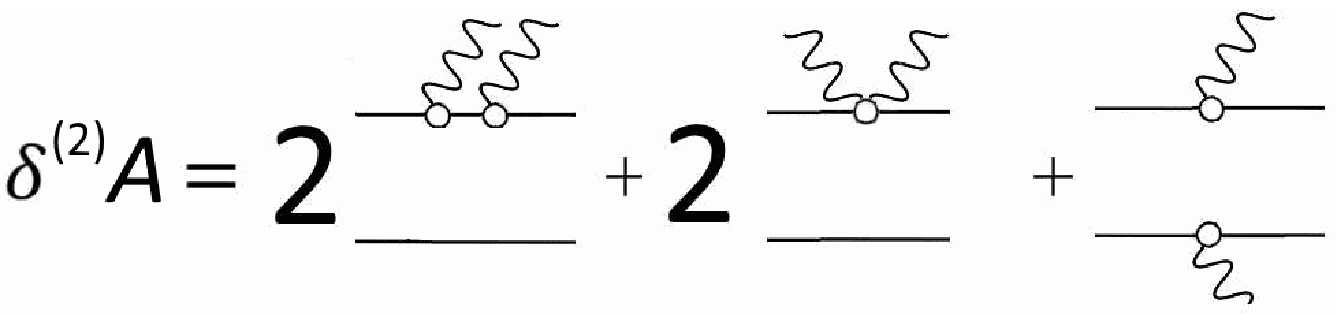}
\caption{ Expression ({\ref{formula11}}) in the diagrammatic representation.}
\label{fig-4}       
\end{figure}

    The quantities $\delta^{(1)}V$ and $\delta^{(2)}V$ are obtained by variation  of Eq.(\ref{formula1}) in the phonon field:
   
\begin{align}
\delta^{(1)}V =& \delta^{(1)}FAV + F\delta^{(1)}AV + FA\delta^{(1)}V, \nonumber \\ 
\delta^{(2)}V =&  \delta^{(1)}\delta^{(1)}V = F\delta^{(2)}AV + \nonumber \\
&2\delta^{(1)}F\delta^{(1)}AV + 
2\delta^{(1)}FA\delta^{(1)}V 
                +2F\delta^{(1)}A\delta^{(1)}V +  \nonumber \\ 
& \delta^{(2)}FAV + FA\delta^{(2)}V  
\label{formula12}
\end{align}

In [I] the quantities $\delta^{(1)}V$ and $\delta^{(2)}V$ were accounted  approximately; namely, only free terms of  equations Eq.(\ref{formula12})  were taken into account for them. This approximation provided  accounting for only 1p1h$\otimes$phonon configurations.

In this article, we reject  this approximation. We transform the obtained equations  Eq.(\ref{formula12})for $\delta^{(1)}V$ and $\delta^{(2)}V$
to the \textit{expressions} for $\delta^{(1)}V$ and $\delta^{(2)}V$ without loss of accuracy. Note that  we use Eq.(\ref{formula4}), i.e. the RPA (or TFFS) approach, for phonons.

Let us rewrite the equations for $\delta^{(1)}V$ and $\delta^{(2)}V$  as follows: 
\bea
\delta^{(1)}V = (\delta^{(1)}V)_0 + FA\delta^{(1)}V, \nonumber \\
\delta^{(2)}V = (\delta^{(2)}V)_0 + FA\delta^{(2)}V,
\label{formula13}
\eea
where $(\delta^{(1)}V)_0$ and  $(\delta^{(2)}V)_0$ are free terms of Eq.(\ref{formula12}). In the other form, we have (symbolically as always):
\bea
(1 - FA)\delta^{(1)}V = (\delta^{(1)}V)_0,   \nonumber \\
(1-FA)\delta^{(2)}V = (\delta^{(2)}V)_0,  
\label{formula14}
\eea
or
\bea
\delta^{(1)}V = (1-FA)^{-1}(\delta^{(1)}V)_0 \nonumber \\
\delta^{(2)}V = (1-FA)^{-1}(\delta^{(2)}V)_0. 
\label{formula15}
\eea

Following  \cite{Khodel}, we introduce also the quantity d$\Gamma$ 
(to avoid mixing with the usual variation of Eq.(\ref{formula3}), we redefined it instead of $\delta \Gamma$ in \cite{Khodel})
 \beq
d\Gamma = \delta^{(1)}F + FAd\Gamma. 
\label{formula16}
\eeq

Further we will use the following symbolic expressions obtained from Eq.(\ref{formula3}) and Eq.(\ref{formula16})
 \bea
\Gamma = (1-FA)^{-1}F \nonumber \\
d\Gamma = (1-FA)^{-1}\delta^{(1)}F  
\label{formula17}
\eea  

Substituting the free terms $(\delta^{(1)}V)_0$ and $(\delta^{(2)}V)_0$ of Eq.(\ref{formula12})
into Eq.(\ref{formula15}) and using Eqs.(\ref{formula17}), we obtain the accurate  expressions for $\delta^{(1)}V$ and  $\delta^{(2)}V$: 
 \bea
\delta^{(1)}V =  d\Gamma AV + \Gamma\delta AV         \nonumber \\
\delta^{(2)}V = \Gamma\delta^{(2)}AV +  2d\Gamma\delta AV + 2d\Gamma A\delta V + \nonumber \\
              2\Gamma\delta A\delta V + d^{(2)}\Gamma AV,
\label{formula18}
\eea
which contain $\Gamma$ and $d\Gamma$ instead of F and $\delta F$
and  we have introduced a new quantity
\bea
d^{(2)}\Gamma = \delta^{(2)}F + FA d^{(2)}\Gamma, 
\label{formula19}
\eea
or
\bea
d^{(2)}\Gamma = (1-FA)^{-1} \delta^{(2)}F. 
\label{formula20}
\eea

The obtained  exact expressions for $\delta^{(1)}V$ and $\delta^{(2)}V$ are shown in Fig.\ref{fig-5}. 
Note, the "accuracy" for $\delta^{(1)}V$ and $\delta^{(2)}V$  consists in the fact that they contain just the TFFS equations for 
vertex V, Eq.(\ref{formula1}), amplitude $\Gamma$, Eq.(\ref{formula3}),  phonon creation amplitude $g$ Eq.(\ref{formula4}) and $d\Gamma$, Eq.(\ref{formula16}), i.e. in this sense everything  is completely within the initial ideas of the standard  TFFS \cite{Migdal}.
The principal difference from [I] is that now we use  the accurate expressions for $\delta^{(1)}V$ and $\delta^{(2)}V$, Fig.\ref{fig-5}, instead of 
free terms of Eq.(\ref{formula12}) for them.

\begin{figure}
\includegraphics[width=1\linewidth]{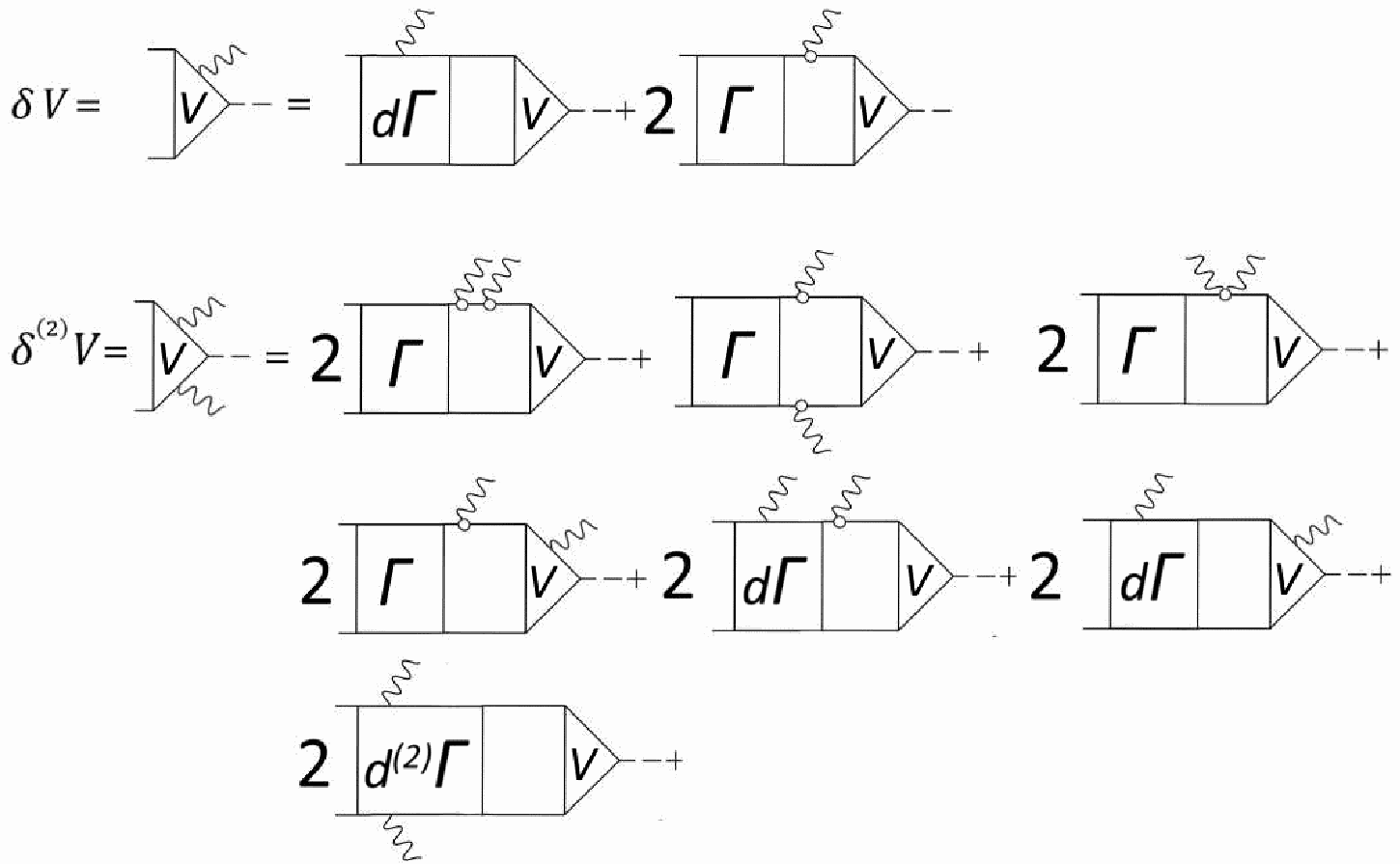}
\caption{Exact expressions, Eq.(\ref{formula18}), for the first and second variations of the vertex V $\delta^{(1)}V$ and $\delta^{(2)}V$ in the phonon field.
The boxes with $\Gamma$, $d\Gamma$ and $d^{(2)}\Gamma$ stand for  quantities  $\Gamma$, Eq.(\ref{formula3}), $d\Gamma$, Eq.(\ref{formula16}) and $d^{(2)}\Gamma$, Eq.(\ref{formula19}), respectively.}
\label{fig-5}       
\end{figure}

\section{The newest equation for the effective field}

\subsection{1p1h$\otimes$phonon configurations  and the full interaction amplitude $\Gamma$}
 Let us go back to  expression (\ref{formula9}). One can see that  expression ({\ref{formula9}) is the first iteration of the following equation (if $V$, Eq.(\ref{formula1}), is zero iteration)       
  \beq
\tilde V = V + \Delta V(g, \tilde V),
\label{formula21}
\eeq  
where $\Delta V(g, \tilde V)$ contains the newest vertex $\tilde V$ in the quantities $\delta^{(1)}V$ and $\delta^{(2)}V$ of Eq.(\ref{formula18}).
Using  Eq.(\ref{formula1}) and Eq.(\ref{formula21}) one can obtain
  \beq
\tilde V = V^0 + FA\tilde V + (1-FA)\Delta V(g, \tilde V).
\label{formula22}
\eeq 
  
Let us substitute into Eq.(\ref{formula22}) the exact expressions for $\delta^{(1)}V$ and $\delta^{(2)}V$, Eq.(\ref{formula18})
(which already contain $\tilde V$)  and use the relations (\ref{formula17}) and (\ref{formula20}). After a long derivation with the use of footnote 1 on page 2, four terms are cancelled and, as the result, the following equation for $\tilde V$ is obtained:

\begin{align}
 \tilde V = e_qV^0 + FA\tilde V +  2FGgDGgGG\tilde V + FGgGDGgG\tilde V+      &\nonumber \\
  2FGg_{\tilde 1\tilde 1}DGG\tilde V + 4gGD\Gamma GgGG\tilde V +                 &\nonumber \\
  2gDGd\Gamma GG\tilde V  + 2\delta FDGgGG\tilde V+                                &\nonumber \\
   2\delta FDGG\Gamma GgGG\tilde V +                                  &\nonumber \\
\delta FDGGd\Gamma GG\tilde V + \delta^{(2)}FGG\tilde V&
\label{formula23}
\end{align}

 This equation contains 10 integral terms instead of 12 in Eq.(16),[I]
 (note that in the analytic form of the equation we write digit 4 in the second line
 of Eq.(\ref{formula23}) instead of digit 2 in graphic representation of two  similar graphs).
 Eq.(\ref{formula23}) can be easily represented in a graphic form. However, for our aim  to include two-phonon configurations 
 (see the next section), it is better to work not with the quantities $d\Gamma$ in Eq.(\ref{formula23}) but with the amplitudes $\Gamma$.
 So, for it, we transform the equation for $d\Gamma$, Eq.(\ref{formula16}), to the\textit{ expression}, which contains  the amplitude $\Gamma$:
  \beq
    d\Gamma = \delta F + \Gamma A\delta F = \delta F + \Gamma GG\delta F.
\label{formula24}
\eeq
Then, substituting Eq.(\ref{formula24}) into Eq.(\ref{formula23}), we obtain the equation for $\tilde V$, which contains only $\delta F$ and
 $\Gamma$:
\begin{align}
\tilde V =  e_qV^0 + FA\tilde V +  2FGgDGgGG\tilde V + FGgGDGgG\tilde V+      &\nonumber \\
  2FGg_{\tilde 1\tilde 1}DGG\tilde V + 4gGD\Gamma GgGG\tilde V +                 &\nonumber \\ 
2gDG\delta FGG\tilde V + 2gDG\Gamma GG\delta FGG\tilde V + 2\delta FDGgGG\tilde V + &\nonumber \\
2\delta FDGG\Gamma GgGG\tilde V +                                                  &\nonumber \\
\delta FDGG\delta FGG\tilde V + \delta F DGG\Gamma GG\delta FGG\tilde V + \delta^{(2)}FGG\tilde V,&
\label{formula25}
\end{align}
It is shown in Fig.\ref{fig-6}. 
.

\begin{figure}
\includegraphics[width=1\linewidth]{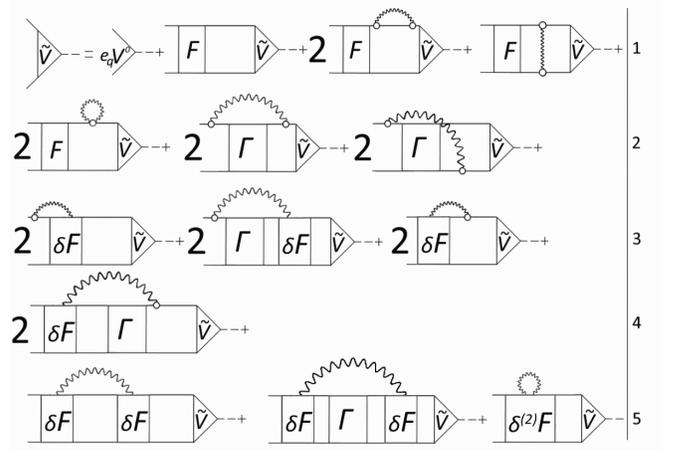}
\caption{Graphic representation of Eq.(25)}
\label{fig-6}       
\end{figure}
Eq.(\ref{formula25}), Fig.\ref{fig-6}, contains only known quantities $\Gamma$, $g$ and $g_{11}$, which satisfy 
Eq.(\ref{formula3}), Eq.(\ref{formula4}) and Eq.(\ref{formula7}), respectively. The quantity $\delta F$ can be estimated with the use of ansatz in Eq.(\ref{formula8}).
Thus, Eq.(\ref{formula25}), Fig.\ref{fig-6}, is   \textit{ the first main result of our article}. It is exact in the sense that  here we have used the exact expressions Eq.(\ref{formula18})
for the first and second variations $\delta^{(1)}V$ and $\delta^{(2)}V$ of the vertex V, Eq.(\ref{formula1}), in  the phonon field.

This is a noticeable generalization of Eq.(16), Fig.6, in  article [I]. Let us compare our Eq.(\ref{formula25}), Fig.\ref{fig-6}, and Eq.(16), Fig.6, of [I].
For simplicity, we enumerate the terms  of Eq.(\ref{formula25}), Fig.\ref{fig-6},  in accordance with their lines as follows: 
 \beq
\tilde V = \tilde V^1 +  \tilde V^{tad} + \tilde V^{2}_n + \tilde V^{3}_n + \tilde V^{4}_n + \tilde V^{5}_n 
\label{formula26}
\eeq 
Here the upper indices mean only the number of the line in  Eq. (\ref{formula25}), Fig.\ref{fig-6}. Some parts of Eq.(\ref{formula26})  may include two or three terms in each line (with digit 2 each). Low index n in four terms of Eq.(\ref{formula26}) shows that these terms contain  new terms as compared to [I].

1. We obtained the full coincidence with [I] in line 1 and for  the first term of line 2 (the dynamical  effect of the tadpole)

2. However, there are  considerable differences. Namely, while Eq.(16), Fig.6, in [I] contains the effective interaction F and $\delta F$, in our case five terms with the full amplitude $\Gamma$ appear.
It gives the possibility to obtain naturally two-phonon configurations (see the  next section). 
 In the terms $ \tilde V^{2}_n$  of line 2 we obtained the terms  similar to [I], but, what is of most interest, in our case they contain
    the full amplitude $\Gamma$, instead of the effective interaction $F$ in the same terms in [I].  $\Gamma$ is not a static quantity
 and  depends on the energy  $\omega$. In this sense, these four  terms are physically similar to the results in \cite{litva-schuck} shown in Fig.13 of \cite{litva-schuck}.
 However, in contrast to \cite{litva-schuck}, here we cannot include the configurations  more complex than two-phonon ones.  
The reason is that in our case it would be incorrect to go  beyond the formulas in Section II.A.

All the terms in lines 3,4,5 of Eq.(\ref{formula25}), Fig.\ref{fig-6}, are new as compared to  TBA  and other models and methods in the PDR and GMR field because they contain $\delta F$, and, among them, there are three terms with  $\Gamma$ in lines 3,4,5.

\subsection{1p1h$\otimes$phonon and two-phonon  configurations}


To obtain all previous results, i.e. to generalize TFFS for the PC case, we have used the initial TFFS formulas  described in 
Section II.A. In order to preserve  the consistency of this approach for the inclusion of two-phonon configurations,
 it is also necessary to use  TFFS.  And Eq.(\ref{formula25}) gives such a possibility. Let us consider the expansion in phonons
for the amplitude $\Gamma$
 \beq
    \Gamma (\omega) =  \sum_s\frac{g^{s}g^{s*}}{\omega -  \omega_s},
\label{formula27}
\eeq
 where $g$ satisfies Eq.(\ref{formula4}). In the PDR and GMR calculations, as a rule, a great number of phonons are used, so that the expansion in  phonons,  
Eq.(\ref{formula27}), exhausts  almost all $\Gamma$.  
 Making use of   the RPA phonons for accounting for PC effects is applied in many modern approaches,
like QPM \cite{Sol89}, TBA and QTBA \cite{ts2007}, relativistic QTBA \cite{litva-ring-ts}. We will see that   our approach for the TFFS generalization  gives some additional effects.

In principle, in Eq.(\ref{formula27}), one could  add a regular part  $\Gamma^r$ that  does not depend on $\omega $.
However, such an approach  will give a strong complication. First, a necessity to find  $\Gamma^r$. Second, one can easily obtain
that the use of  $\Gamma^r$ will strongly complicate  the equation for $\tilde V$. Indeed, in this case, the quantity  $\Gamma^r$ will appear in line 
2 of Fig.\ref{fig-6} and in other lines containing  $\delta F$. These new terms will generalize the results of [I] in such a way that the new quantities $\Gamma^r_{ind}$, which differ from $F_{ind}$ in [I] only by F  changed by an  unknown quantity $\Gamma^r$, should be in 
line 2 of Fig.\ref{fig-6} and in other lines. Such a complication is not constructive  at this stage and will be discussed somewhere else.
For these reasons, in this article we  omit $\Gamma^r$.
Substituting  Eq.(\ref{formula27}) into the term $2gDGg\Gamma G$, which plays the role of a phonon-induced interaction, we have (symbolically): 
\beq
gDG\Gamma Gg =  gGgDDgGg,
\label{formula29}
\eeq
which is shown  in a graphic form in Fig.\ref{fig-7}. However, note that, for example, for the problems with consideration  of
two specific  phonons, in principle, it will be necessary to find $\Gamma^r$.   

\begin{figure}
\includegraphics[width=1\linewidth]{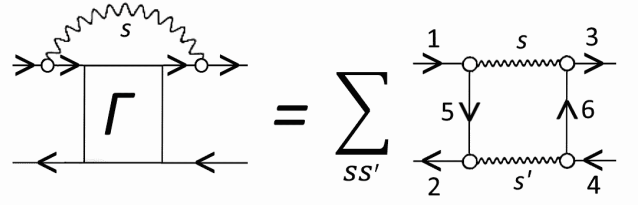}
\caption{Graphic representation of Eq.(\ref{formula29}).}
\label{fig-7}       
\end{figure}

In order to obtain the final newest equation for the vertex $\tilde V$, it is necessary to substitute 
Eq.(\ref{formula27})  into all five terms of Eq.(\ref{formula25}), Fig.\ref{fig-6}, which contain the amplitude $\Gamma$. The result is as follows:

\begin{align}
 \tilde V = e_{q}V^0 + FA\tilde V + 2FGgDGgGG\tilde V + FGgGDGgG\tilde V& +\nonumber \\ 
             2FGg_{\tilde 1\tilde 1}DGG\tilde V + & \nonumber \\
              4gGgDDgGgGG\tilde V +&                              \nonumber \\ 
            2g\delta FDGGG\tilde V + 2\delta FDGgGG\tilde V +&   \nonumber \\ 
                 2\delta FDGG\delta FGG\tilde V +  \delta^{(2)}FDGG\tilde V +& \nonumber \\ 
                 \delta FGGgDDgGG\delta FGG\tilde V +& \nonumber \\
              2gGgDDgGG\delta FGG\tilde V +     & \nonumber \\
              2\delta FGGgDDgGgGG\tilde V  &                                                                            
 \label{formula30}
\end{align}
 It is shown in Fig.\ref{fig-8}.   
 The lines of Eq.({\ref{formula30}) and Fig.\ref{fig-8} correspond to each other. Due to the symbolical  form  of Eq.({\ref{formula30}), in
   line 3, we have shown two graphs in Fig.8 instead of the terms with  digit  4 in the same   line 3 of Eq.({\ref{formula30})

\begin{figure}
\includegraphics[width=1\linewidth]{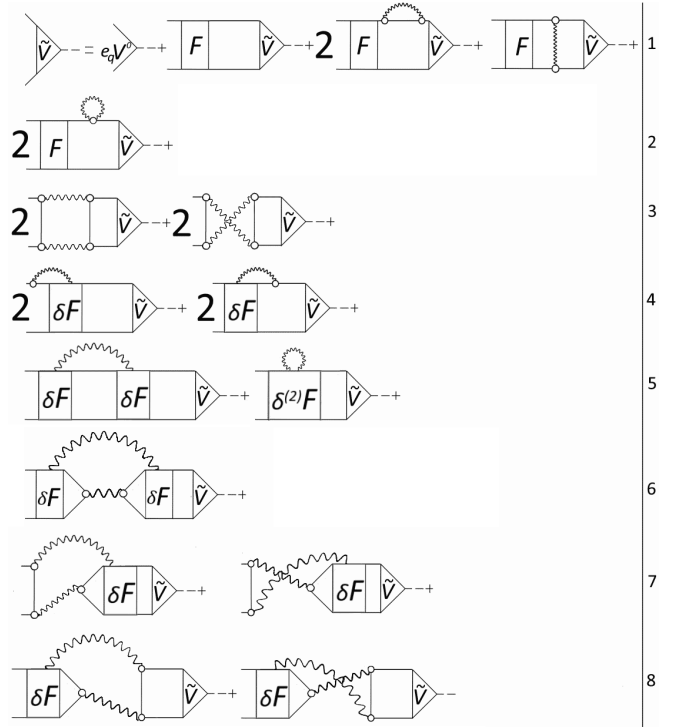}
\caption{Graphic representation for Eq.(\ref{formula30})}
\label{fig-8}       
\end{figure}

Using the standard diagram technique, one can easily see that Eq.({\ref{formula30}), Fig.\ref{fig-8}, contains both 1p1h$\otimes$phonon  configurations in all the lines, except for the RPA part (Eq.(\ref{formula1})) in line 1, and two-phonon  configurations
in lines 3, 6, 7, 8.  Thus, Eq.(\ref{formula30}), Fig.\ref{fig-8}, is the \textit{second main result of our article}.

\section{Discussion of the newest equation for the effective field $\tilde V$}

\subsection{ General description. \\Comparison to  article [I] } 

In subsections A and B of the previous Section IV, Fig.\ref{fig-6} and Fig.\ref{fig-8}, respectively,  we obtained the justification of our ansatz in [I], where the very first integral term $FA\tilde V$ was included intuitively
instead of $FAV$. This justification is due to inclusion of the exact expressions for $\delta^{(1)}V$ and $\delta^{(2)}V$.

Like in [I] and also in the case of 1p1h$\otimes$phonon configuration in the previous section, see Eq.(\ref{formula25}), Fig.\ref{fig-6}, we obtained the full coincidence with [I] in line 1 and in the first term of line 2 (the dynamical tadpole effect).
 Besides, this is  a considerable generalization of Eq.(16), Fig.6, in  article [I].
As compared to  Eq.(16), Fig.6 of [I], we obtained  new five terms with two-phonon configurations in lines 3, 6, 7, 8 
 in Eq.(\ref{formula30}).

All the terms in lines 4, 5, 6, 7, 8 of Eq.(\ref{formula30}), Fig.\ref{fig-8}, are quite new as compared to  TBA  and other models and methods  in the PDR and GMR field because they contain $\delta F$.   

For simplicity, we enumerate the terms  of Eq.(\ref{formula30}), Fig.\ref{fig-8},  in accordance with their lines as follows:
 \bea
\tilde V = \tilde V^1 +  \tilde V^{tad} +  \tilde V^{3}_{2phon} \nonumber \\
 + \tilde V^{4} + \tilde V^{5} +  + \tilde V^{6}_{2phon} + \tilde V^{7}_{2phon} + \tilde V^{8}_{2phon}
\label{formula31}
\eea 
Here the upper indices 1-8 mean only the number of the line in  Eq.(\ref{formula30}), Fig.\ref{fig-8}. The low indices $2phon$ mean that  the terms contain  two-phonon configurations. Some parts of Eq.(\ref{formula31})  may include two terms in each line.

1. We obtained the full coincidence between Eq.(16), Fig.6[I], and  Eq.(\ref{formula30}), Fig.\ref{fig-8}, in lines 1 and   lines 2 (the  dynamical   effect of the tadpole).
 See the discussion about these terms in Section 4 in [I].
 
2. Four  terms  in lines 4 and 5, $\tilde V^{4}$ and  $\tilde V^{5}$, coincide with $\tilde V^{4}$ and  $\tilde V^{6}$ 
in [I], respectively, they were  obtained and discussed there. They are of order $g^2$. 

3. The terms in lines 6, 7, 8, i.e. $\tilde V^{6}_{2phon} , \tilde V^{7}_{2phon}$ and $\tilde V^{8}_{2phon}$, contain 
two-phonon configurations and $(\delta F)^{2}g^2$, $(g^3\delta F)$ and $(\delta F)g^3$,  respectively. Since $\delta F$
contains g, see Eq.(\ref{formula8}), all these terms are of order $g^4$.

All the terms in lines 4-8 contain the quantity $\delta F$ , $(\delta F)^2$ or $\delta^{(2)} F$.
The quantity $\delta F$ is expressed in terms of the three-quasiparticle   effective interaction amplitude $W$ \cite{KhSap1982}:
\beq
\delta_{s}F = WGg_{s}G.
\label{32}
\eeq
As the role of this interaction is known to be small on the whole,
one can think that the terms with $\delta F$ will give  a small contribution. We have  only one real example when it was estimated numerically
  \cite{voitenkov} with the use of  Eq.(\ref{formula8}): the  contribution of  the term with $\delta F$ was very small for the case of static characteristics.    
So, in the present article, we will not consider these lines  and further down we will  consider  only terms  $\tilde V^{3}_{2phon}$, see Section B.  We will obtain general formulas for them, but, first of all, and in a more detailed form, we consider quite new phonon-induced interactions that  appear in line 3, which  are caused by the  phonon exchange in various channels (ph, hp, pp, hh).
 
 \subsection{Terms $\tilde V^3_{2phonon}$ (line 3).  Two-phonon configurations. \\ Comparison to the TBA model.}

  Here we  introduce the two-phonon induced interaction $F^{2phonon}_{ind}$ for the first of the two two-phonon graphs of Fig.\ref{fig-8}:

\bea
(F^{2phonon}_{ind})_{1234\_1}(\eps_1,\eps_3,\omega) = \nonumber\\
\sum_{56ss'}g^s_{15}g^{s*}_{63}g^{s'}_{52}g^{s'*}_{46}I_{56ss'\_1}\delta(\eps_1-\eps_2+\eps_4-\eps_3),
\label{formula48}
\eea
\bea
I_{56ss'\_1}(\eps_1,\eps_3)=\nonumber\\
\int G_5(\eps_1-\omega_1)G_6(\eps_3-\omega_1)D_s(\omega_1)D_{s'}(\omega_1-\omega)d\omega_1.
\label{formula49}
\eea

The first  two-phonon induced interaction $F^{2phonon}_{ind}$, which  is present in   the second ("crossed") graph of line 3, Fig.(\ref{fig-8}), reads:  

\bea
(F^{2phonon}_{ind})_{1234\_2}(\eps_1,\eps_4,\omega) = \nonumber \\
\sum_{56ss'}g^s_{15}g^{s*}_{46}g^{s'}_{52}g^{s'*}_{63}I_{56ss'\_2}\delta(\eps_1-\eps_2+\eps_4-\eps_3),
\label{formula50}
\eea
\bea
I_{56ss'\_2}(\eps_1,\eps_4)=\nonumber\\
\int G_5(\eps_1-\omega_1)G_6(\eps_4+\omega_1)D_s(\omega_1)D_{s'}(\omega_1-\omega)d\omega_1.
\label{formula51}
\eea

The results of integration in Eq.(\ref{formula49}) and Eq.(\ref{formula51}) are  given in Eq.(\ref{formula52}), where we introduced 
$(\eps_1-\eps_2) = (\eps_3-\eps_4) = \omega$. 
Eq.(\ref{formula52}) was obtained with the help of computer transformations from the initial much more cumbersome results of integrations in order to try and single out (unsuccessfully !) only terms with $[\omega \pm (\omega_s + \omega_s^{\prime})]^{-1}$. 

\begin{table*}[t!]
\begin{align}
&I_{56ss'\_1}(\eps_1,\eps_3,\omega)=\nonumber\\
&\frac{1-n_{\lambda_5}}{(\eps_3+\eps_{\lambda_5}-\eps_{\lambda_6}-\eps_1)(\eps_1-\eps_{\lambda_5}-\omega_s)(\eps_1-\eps_{\lambda_5}-\omega-\omega_s')}-\frac{n_{\lambda_5}}{(\eps_3+\eps_{\lambda_5}-\eps_{\lambda_6}-\eps_1)(\eps_1-\eps_{\lambda_5}+\omega_s)(\eps_1-\eps_{\lambda_5}-\omega+\omega_s')}+\nonumber\\
&\frac{(1-n_{\lambda_5})n_6}{(\eps_3+\eps_{\lambda_5}-\eps_{\lambda_6}-\eps_1)(\eps_3-\eps_{\lambda_6}+\omega_s)(\eps_3-\eps_{\lambda_6}-\omega+\omega_s')}-\frac{n_{\lambda_5}(1-n_6)}{(\eps_3+\eps_{\lambda_5}-\eps_{\lambda_6}-\eps_1)(\eps_3-\eps_{\lambda_6}-\omega_s)(\eps_3-\eps_{\lambda_6}-\omega-\omega_s')}+\nonumber\\
&\frac{1}{(\eps_1-\eps_{\lambda_5}+\omega_s)(\eps_3-\eps_{\lambda_6}+\omega_s)(\omega_s+\omega_s'+\omega)}+\frac{1}{(\eps_1-\eps_{\lambda_5}-\omega_s)(\eps_3-\eps_{\lambda_6}-\omega_s)(\omega_s+\omega_s'-\omega)}\
\label{formula52}
\end{align}
\begin{align}
&I_{56ss'\_2}(\eps_1,\eps_4, \omega)= \nonumber\\
&\frac{1-n_{\lambda_5}}{(\eps_4+\eps_1-\eps_{\lambda_6}-\eps_{\lambda_5})(\eps_1-\eps_{\lambda_5}-\omega_s)(\eps_1-\eps_{\lambda_5}-\omega-\omega_s')}-\frac{n_{\lambda_5}}{(\eps_4+\eps_1-\eps_{\lambda_6}-\eps_{\lambda_5})(\eps_1-\eps_{\lambda_5}+\omega_s)(\eps_1-\eps_{\lambda_5}-\omega+\omega_s')}-\nonumber\\
&\frac{(1-n_{\lambda_5})(1-n_6)}{(\eps_4+\eps_1-\eps_{\lambda_6}-\eps_{\lambda_5})(\eps_{\lambda_6}-\eps_4+\omega_s)(\eps_{\lambda_6}-\eps_4-\omega+\omega_s')}+\frac{n_{\lambda_5}n_6}{(\eps_4+\eps_1-\eps_{\lambda_6}-\eps_{\lambda_5})(\eps_{\lambda_6}-\eps_4-\omega_s)(\eps_{\lambda_6}-\eps_4-\omega-\omega_s')}+\nonumber\\
\nonumber\\ 
&\frac{1}{(\eps_1-\eps_{\lambda_5}+\omega_s)(\eps_4-\eps_{\lambda_6}-\omega_s)(\omega_s+\omega_s'+\omega)}+\frac{1}{(\eps_1-\eps_{\lambda_5}-\omega_s)(\eps_4-\eps_{\lambda_6}+\omega_s)(\omega_s+\omega_s'-\omega)}\nonumber
\end{align}

\end{table*}

In Fig.\ref{fig-10}, we show two-phonon induced interactions $(F^{2phonon}_{ind})_{1234\_1}$, Eq.(\ref{formula48}) and $(F^{2phonon}_{ind})_{1234\_2}$
 Eq.(\ref{formula50}), which are present in line 3 of Eq.(\ref{formula30}), Fig.(\ref{fig-8}), and illustrate  
  possible  1p1h$\otimes$phonon and
two-phonon configurations created due to accounting for ground state correlations (GSC), or "graphs going back".
These configurations may be clearly seen from the cross-cuts shown  by dashed lines.  They correspond to numerous 
denominators in Eq.(\ref{formula52}).
From the initial much more cumbersome results of integrations mentioned above, 
one can see more clearly that the two-phonon denominators 
$[\omega \pm (\omega_s + \omega_{s^{\prime}}]^{-1}$ are present in both  pp($n_{\lambda_5}n_{\lambda_6}$), hh$(1-n_{\lambda_5})(1-n_{\lambda_6})$
and hp $(1-n_{\lambda_5})(n_{\lambda_6})$, ph$(n_{\lambda_5})(1-n_{\lambda_6})$ terms, where pp (hh) corresponds to two particles  (holes)  above  (below)  
the   Fermi surface and hp( ph) corresponds to a hole and a particle  which lie on different sides of the Fermi surface.
Note, the two-phonon terms in line 3 (see Fig.\ref{fig-10} and Eq.(\ref{formula52}))   contain  complex 1p1h$\otimes$phonon and two-phonon configurations.
Our two-phonon configurations also contain  two-phonon GSCs with the denominators 
$[\omega + (\omega_s + \omega_{s^{\prime}}]^{-1}$. 
Thus, in addition to line 1, 1p1h$\otimes$phonon configurations are present  in  line 3, i.e. we get a  considerable complication 
 as compared to [I].
\begin{figure}[h]
\includegraphics[width=1\linewidth]{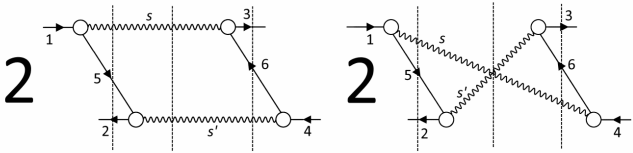}
\caption{Two-phonon induced interaction $(F^{2phonon}_{ind})_{1234\_1}$, Eq.(\ref{formula48}) and $(F^{2phonon}_{ind})_{1234\_2}$
 Eq.(\ref{formula50}), which are present  in line 3 of Eq.(\ref{formula30}),Fig.(\ref{fig-8}). Illustration of GSCs for  two-phonon graphs
in line 3 of Fig.(\ref{fig-8}). The
cross-cuts show various 1p1h$\otimes$phonon and two-phonon configurations.}
\label{fig-10}       
\end{figure} 

 Here one can see a considerable difference from the two-phonon version of (Q)TBA \cite{ts2007,litva-ring-ts}. Our method of introduction of two-phonon configurations  shown in Eq.(\ref{formula29}), Fig.7, gives   much more complicated $\omega$-dependence for our two-phonon induced interactions $F^{2phonon}_{ind}$, Eq.(\ref{formula48}) and Eq.(\ref{formula50}), which correspond to the $\omega$-dependent two-
quasiparticle amplitude $\Phi (\omega)$ in \cite{ts2007,litva-ring-ts}.  The difference  is due to the use of a special factorization procedure for
 $\Phi (\omega)$ in \cite{ts2007,litva-ring-ts}. Namely, it consists in 
inclusion of the correlations in the ph-pair entering
the 1p1h$\otimes$phonon configuration, i.e., replacement of the
uncorrelated pair with  the phonon  \cite{ts2007}. We think that our method with the full 1p1h-interaction amplitude $\Gamma$ shown in Eq.(\ref{formula29}), Fig.7, is rather natural.

 It is necessary to take into account the general specific features  of the self-consistent approach. We mean the subtraction method developed in detail  
 by V. Tselyaev \cite{tssubtraction,litva-ring-ts}. For the non-self-consistent approach, the analogous procedure was described and realized in \cite{kaev83} and \cite{ktZPhys},
 respectively, within the so-called refinement procedure that  included  the refined   single-particle basis. As known, the energy density functional is constructed 
 in the way to give the exact (in the limiting case) description of the nuclear ground state properties. Therefore, to avoid double counting of
 static contributions of complex configurations, which are already contained in the energy density functional, in our case  it is necessary to subtract the static part 
 $F^{2phonon}_{ind} (\omega =0)$ from   $F^{2phonon}_{ind} (\omega )$. The detailed discussion of these rather  numerous problems
 goes beyond the scope of our work.

The terms $\tilde V^3_{2phonon}$ in line 3 of Eq.(\ref{formula30}), Fig.\ref{fig-8}, can be written through our two-phonon induced interactions, 
Eq.(\ref{formula48}) and Eq.(\ref{formula50}):

 \begin{align}
(\tilde V^3_{2phon})_{12\_1}(\eps_1,\omega) = \nonumber \\
\sum_{34}\int (F^{2phonon}_{ind})_{1234\_1}(\eps_1,\eps_3)G_3(\eps_3+\omega)G_4(\eps_3)\tilde V_{34} d\eps_3 \nonumber \\
(\tilde V^3_{2phon})_{12\_2}(\eps_1,\omega) = \nonumber \\
\sum_{34}\int (F^{2phonon}_{ind})_{1234\_2}(\eps_1,\eps_4)G_3(\eps_4+\omega)G_4(\eps_4)\tilde V_{34} d\eps_4 
\label{formula53} 
 \end{align}

The final expressions for  $(\tilde V^{3}_{2phon})_{12}(\eps_1,\omega)$ are rather cumbersome.
They will be obtained and discussed in details somewhere else.

\section{Energies and probabilities of transitions}
Formulas for the energies and probabilities of transitions between the ground and excited states were  obtained in [I] by means of 
direct and rather formal generalization of the respective formulas in  TFFS \cite{Migdal, Migdal2}. For the simplest case, which corresponds to the terms of $\tilde V^1$, Eq.(\ref{formula31}) or lines 1 of Fig.\ref{fig-8} and  Fig.\ref{fig-6}, they were realized numerically without the TBA prescription in  articles \cite{ktZPhys,kt1984} for fine structure characteristics  of 
 isovector M1 resonances in magic nuclei. With the use of the TBA prescription and within the strongly improved TBA version, they were realized 
for the  M1 resonance in  $^{208}Pb$ recently in  articles  \cite{Tsel2019,Tsel2020} (see   modifications of the TBA model there). In all
these calculations there were no full explanation of the fine structure of  the M1 resonance in $^{208}Pb$.

In [I] and in the present article, the next important step is realized, namely, due to inclusion of the first term of Eq.(\ref{formula10}), Fig.\ref{fig-3}, 
the new phonon-induced interactions  appear and, therefore,  the $\eps_1$-dependence of $\tilde V(\eps_1)$ is obtained for the first time.
 \footnote{ In the present article, the $g^4$ effects
are already considered in the two-phonon graphs. So the arguments in [I] about $\rho = \int GG\tilde Vd\eps$ in the $g^2$
approximation  are not suitable in our case.   However, it  does not mean that  the $g^4$ effects are not important. They are important just for the fine structure where they 
will result in  the  strength redistribution.} 

This dependence, although  interesting as it is, must be  integrated  and, therefore, is not seen  in the strength function 
\beq
S(\omega, \Delta) = \frac{dB(EL)}{d\omega} = -\frac{1}{\pi}Im\sum_{12}e_{q}V^{0}_{12}\rho_{21}(\omega + \imath\Delta), 
 \label{formula53}
 \eeq
 where the density matrix $\rho = \overline{A}\tilde V$, $\overline A$ is a new generalized propagator and 
 $\Delta$ is a smearing parameter that simulates a finite experimental resolution. 
 
 In our case, the equation for 
 the density matrix $\rho$  is  obtained  by means of  generalization of the equation for $\rho$ in [I]. 
 From Eq.(\ref{formula53}),  one can  obtain the transition probabilities and energy-weighed sum rule summed over an energy interval. 
 This prescription  is similar to the method of strength functions  always used  in the QPM method. As far as the important  fine structure problem 
 is concerned, the fine structure characteristics can be obtained at small numbers   $\Delta$ = 10 keV or 1 keV. In \cite{Lyutor2018} such 
 calculations for the PDR in $^{208}Pb$ were performed within the self-consistent TBA with the use of Skyrme forces, and they   showed that
 it was   impossible to obtain a reasonable agreement with the observable PDR fine structure. One can hope that the calculations within our approach should improve the situation.
 
\section{Conclusion}
In this article, the self-consistent TFFS has been  consistently generalized  for  the case of accounting for PC effects 
in  the  energy region of PDR and GMR   with the aim  to take into account both 1p1h$\otimes$phonon and two-phonon configurations.

 If  to neglect  the terms with $\delta F$, which probably give small contributions,    our new results 
are  contained in two variants of the newest equation for  vertex $\tilde V$  shown in Fig.6 and Fig.8.
The  first variant, Section IV.A, Fig.6, contains 1p1h$\otimes$phonon configurations and the full amplitude 1p1h-interaction $\Gamma$, 
Eq.(\ref{formula3}). This variant has some promising prospects
 for further development.

 In the second variant, Section IV.B, Fig.\ref{fig-8}, the first step for further development has been realized
within our TFFS generalization   for the vertex $\tilde V$.  Namely,  through expansion of the full amplitude interaction  $\Gamma$
in the RPA phonons, Eq.(\ref{formula27}),
it was possible to add  two-phonon configurations  
 to  1p1h$\otimes$phonon ones  in  line 3 (and also in lines 6,7,8 of Eq.(\ref{formula30}), Fig.\ref{fig-8},
containing  terms with $\delta F$) and to obtain the new two-phonon-induced  interaction $F^{2phonon}_{ind}$.  

 In this second variant, both 1p1h$\otimes$phonon and two-phonon configurations, whose necessity was discussed in the Introduction, have been obtained. This allowed us to compare our results  with the known  QPM and QTBA  approaches (in their variants for magic nuclei). We obtained a considerable complication as compared to them and  article [1], not to mention 
the terms with $\delta F$. Our
 two-phonon part  strongly differs from the TBA two-phonon part: it contains a more complex $\omega$-dependence, i.e.  both two-phonon configurations with the numerators $[\omega \pm (\omega_{s^{\prime}} + \omega_s)]^{-1}$ and 1p1h$\otimes$phonon configurations  together, see Eq.(\ref{formula52}).

 In both variants we confirmed the previous model \cite{kaev83} shown in Fig.\ref{fig-1} and, in fact, the standard
TBA model \cite{ts89} but only as a particular  case corresponding to  line 1 in both variants shown in Fig.\ref{fig-6}  and Fig.\ref{fig-8} (of course, provided that in the terms of line 1 and in other lines, the MCDD, or TBA prescription, 
will be applied, if necessary). Like in the considered case of [I], in both variants, the dynamic term with $\tilde V^{tad}$ is present. In 
the  PDR+GMR energy region , this is a quite new effect and it is necessary to solve
Eq.(\ref{formula7}) for the two phonon creation amplitude $g_{11}$, in order to consider it. It can be done in the representation of single particle wave functions,
but nobody has done it yet.

The two-phonon-induced interactions  $F^{2phonon}_{ind}$  are caused by the  phonon exchange in various channels (ph, hp, pp, hh). 
 Their dependence on energy variables    may be   of interest both for nuclei and, probably, for other Fermi-systems and should be studied in future.

All the obtained formulas contain numerous GSCs in both 1p1h$\otimes$phonon and two-phonon configurations. These effects were investigated 
 mainly within the GF formalism, however, it was not sufficient. They should  be important, at least, in the problem of the PDR and GMR resonance fine structure, see  recent article \cite{nester2020} and references in it.   

Thus, in the present  article, an important step to the direction of consistent inclusion of PC effects to the self-consistent TFFS has been made.
In the nearest future we will finalize   the general  formulation of this approach in  the PDR and GMR energy region with the aim to account for  complex 1p1h$\otimes$phonon and two-phonon configurations. But at the present stage one can already see  that the situation is rather complicated. Suffice it to compare the old approach shown in Fig.\ref{fig-1} and the new one in Fig.\ref{fig-8}. We hope that several aspects of the particle-vibration coupling scheme have been  clarified. Numerous future calculations must clarify the role of new considered effects. A further development is probably  possible  if to go beyond the main initial TFFS equations mentioned  in Section II.A.

\section{ ACKNOWLEDGMENTS}

We are grateful to
V.A. Khodel, and
 V.I. Tselayev  for  useful discussions. S.K. thanks Dr. A.C. Larsen and the Oslo group for stimulating cooperation in the PDR field.
The reported study was funded by RFBR, project no.19-31-90186
 and supported by the Russian Science Foundation, project no.16-12-10155.


\end{document}